\documentclass[a4paper]{article}
\pdfoutput=1

\usepackage[utf8]{inputenc}
\usepackage[T1]{fontenc}
\usepackage[english]{babel}

\usepackage[pdftex,breaklinks=true,bookmarks=true,pdfborder={0 0 0},colorlinks=false]{hyperref}

\usepackage{listings}
\usepackage{amsfonts}
\usepackage{amsmath}
\usepackage{cleveref}
\usepackage{graphicx}
\usepackage[caption=false]{subfig}

\newcommand{\ie}{i.e.~}

\newcommand{\vs}{vs.~}

\usepackage{enumitem}
\setlist{nolistsep}
\usepackage{array}

\begin{document}

\author{Raphael ‘kena’ Poss\\University of Amsterdam, The Netherlands}
\title{The essence of component-based design\\and coordination}

\maketitle

\begin{abstract}
  Is there a characteristic of coordination languages that makes them
  qualitatively different from general programming languages and
  deserves special academic attention?  This report proposes a nuanced
  answer in three parts. The first part highlights that coordination
  languages are the means by which composite software applications can
  be specified using components that are only available separately, or
  later in time, via standard interfacing mechanisms. The second part
  highlights that most currently used languages provide mechanisms to
  use externally provided components, and thus exhibit some elements
  of coordination. However not all do, and the availability of an
  external interface thus forms an objective and qualitative criterion
  that distinguishes coordination.  The third part argues that despite
  the qualitative difference, the segregation of academic attention
  away from general language design and implementation has non-obvious cost
  trade-offs.
\end{abstract}

%\clearpage

\setcounter{tocdepth}{1}
\tableofcontents

\clearpage

\section{Motivation}

During the period 2012-2013 the author was spectator to the evolution
of S-Net~\cite{shafarenko.09.apc,penczek.10.snet,grelck.10.ijpp}, a programming language for streaming networks. To this
author, a characteristic feature of the research activities around
S-Net is the insistence that ``S-Net is a coordination language,''
which pitches S-Net implicitly against ``regular programming
languages'' and suggests that programming for coordination is somehow different from
programming for computation.

Meanwhile, the EU-funded project ADVANCE has been supporting effort to
optimize the execution of S-Net programs on multi-core
computers. During this effort, it has become apparent that S-Net
suffers from similar technical issues as the ``regular programming
languages'' it is usually pitched against, namely in two areas: time
and space scheduling, and memory allocation. In particular, ADVANCE
has revealed a striking overlap of issues and their potential
solutions between S-Net and Single-Assignment C, a functional language
developed by the same researchers.

This author estimates that this convergence of research issues has
caused an \emph{identity crisis} of sorts around S-Net. The crisis
was especially visible during a technical meeting in early June 2013,
where Alex Shafarenko, designer of S-Net, was failing to convince Kath
Knobe, designer of Intel's Concurrent Collections, about what makes
coordination fundamentally special.

The crux of the argument was to determine whether the existence of a
mechanism to define ``black boxes'' in a language is a clear criterion
that separates coordination languages from other languages. The
counter argument was that most languages, including C,
Single-Assignment C, Haskell and all those mentioned during the
discussion, also enable a programmer to define black boxes at any level of
abstraction. By this counter argument, all these language are also
coordination languages as well.

The reason why this discussion matters is that the existence of a
criterion to identify coordination is a prerequisite to motivate
research specialized in ``coordination languages and systems'' and
justify a specialized branch of research and expertise, separate from
general programming language design and implementation. Without this
conceptual frontier, there would be little remaining justification to
continue further effort in developing S-Net and its derived
technologies, or requesting funding to that effect.

In this particular discussion, the participating individuals eventually agreed
that they have observed \emph{perceived merit} in their work from their
community. From then, they were able to conclude that their work must
be worthy of further effort, even though they could not clearly
\emph{express} why at the time.

This situation was uncomfortable to this author from a conceptual
perspective, and this discomfort motivated the production of the
present technical report. In the following sections, we propose
a formulation of the essence of component-based design (\cref{sec:cbd}), including
the distinction between component specification and
instantiation (\cref{sec:specins}), and the essence of coordination
(\cref{sec:coord}). We then discuss the trade-offs of specializing
research towards coordination systems in \cref{sec:discuss}.

\section{Component-based design}
\label{sec:cbd}

The word ``component'' is both versatile and usually well-understood.
A simple definition can be found in~\cite{batory.92}: components are defined by
their \emph{interface}, which specifies how they can be used
in applications, and one or more \emph{implementations} which define
their actual behavior.

The two general principles of \emph{component-based design} are then
phrased as follows. The first is \emph{interface-based integration}:
when a designer uses a component for an application, he agrees to only
assume what is guaranteed from the interface, so that another
implementation can be substituted if needed without changing the rest
of the application. The second is \emph{reusability}: once a component
is implemented, a designer can reuse the component in multiple
applications without changing the component itself.

Component-based design is embedded in different programming paradigms
using different abstractions. For example, in object-oriented
languages, classes define components: the set of methods defines the
component interface, and the set of attributes and method
implementations define the component implementation. In functional
languages, individual functions can be seen as components: the
function signature (list of argument and return types) define its
interface, whereas the function definition (``right-hand side'')
defines its implementation. Other examples are given in
\cref{tab:compspec}.

\section{Component specifications and instances}
\label{sec:specins}

\begin{table}
\centering
\begin{tabular}{p{.21\textwidth}p{.35\textwidth}p{.35\textwidth}}
Abstraction & How interfaces are defined & How implementations
are defined \\
\hline
Classes (OOP) & Method interface & Method code and attributes
\\
Functions (FP) & Function signature & Function
code \\
Unix commands & Manual page (list of command-line arguments and
program description) & Executable file \\
Network service & Protocol & Service implementation \\
Hardware & Signalling specification & Logic design \\
\end{tabular}
\caption{How components are defined in different paradigms}\label{tab:compspec}
\end{table}

We also need to acknowledge a further distinction which is less commonly
found discussed: the difference between
\emph{component specification} and \emph{component instance}.

\begin{figure}
\centering
\subfloat[Specification of a web
CRM.]{\includegraphics[scale=.4]{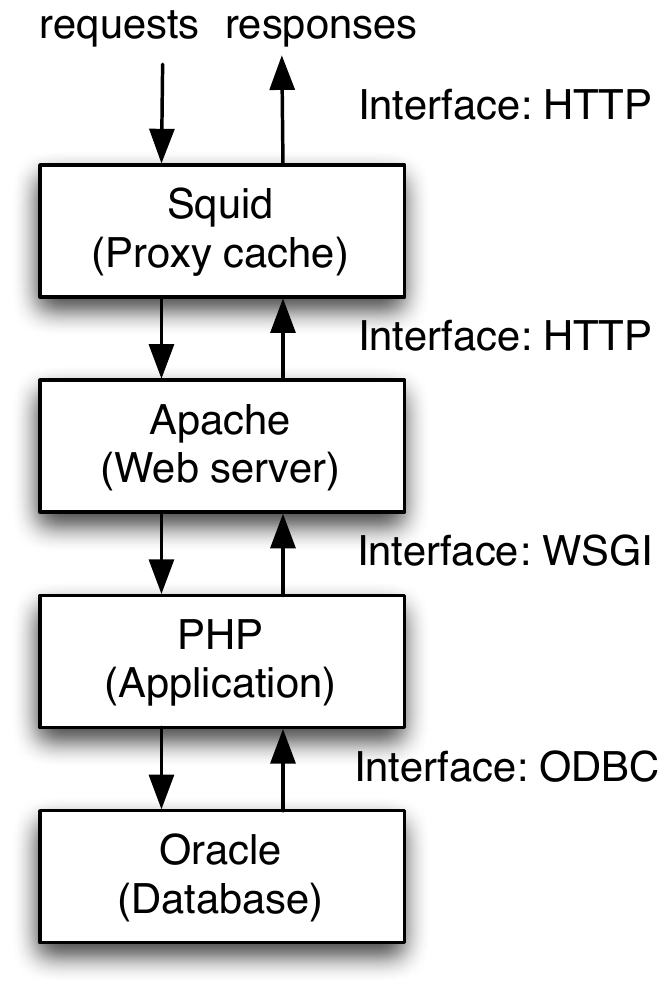}\label{fig:webspec}}
\quad \quad \subfloat[A particular instance of the web
CRM.]{\includegraphics[scale=.4]{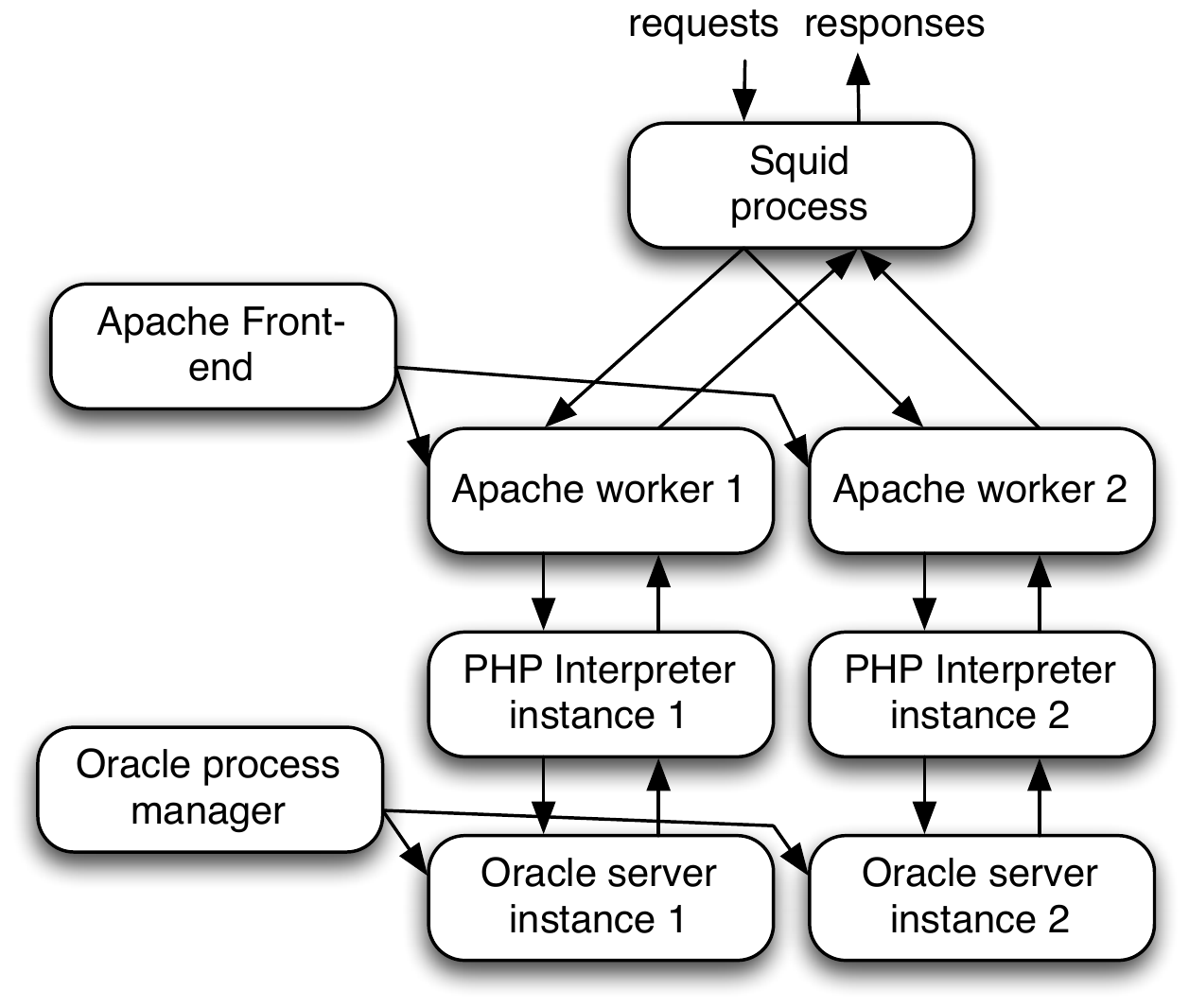}\label{fig:webinst}}
\caption{Specification and instance of a web CRM.}\label{fig:webcrm}
\end{figure}

To illustrate our distinction, we can consider the perspective of a
software engineer tasked with designing a web CRM, who decides to
realize the work by combining a proxy cache, a web server, PHP and a
database server. From this engineer's perspective, the ``advertised''
structure of the application is likely to conform to
\cref{fig:webspec}, which highlights the logical relationship between
the 4 components the engineer has reused. In contrast, the system
administrator who observes the application at run-time may instead
observe the situation described in \cref{fig:webinst}. Here, contrary
to the ``abstract'' specification in \cref{fig:webspec}, the Squid
proxy process does not communicate with the Apache server directly;
instead it communicates with two worker instances spawned by the
Apache server. Each worker instance has in turn spawned its own PHP process
to process its incoming requests. On the database side, a
duplication has also occurred: each time a PHP process requests an ODBC
connection, Oracle creates a new server process specific to that
connection. In this run-time scenario, 9 components are involved
instead of 4.

In the rest of this discussion we name \emph{component specification}
the result of the design work by the programmer, and \emph{component
  instance} the real-world representation of a specification at
run-time. As the previous example shows, each instance is indirectly
``caused by'' one specification, but a single specification may
``cause'' multiple instances. Again, the distinction between
specification
and instance is found in many shapes across computing domains; related
terms are given in \cref{tab:modinst}.

\begin{table}
\centering
\begin{tabular}{lcc}
Conceptual domain & Word for blueprint & Word for real-world
reification \\
\hline
Object-oriented programming & ``class'' & ``object'' \\
Functional programming & ``function'' & ``activation record'' \\
Operating systems & ``program'' & ``process'' \\
Software builds & ``source code'' & ``object code'' \\
Instruction execution & ``executable code'' & ``instruction stream''
\\
Computer architecture & ``design'' & ``implementation'' \\
Simulation & ``model'' & ``simulator'' \\
Parsers & ``grammar'' & ``parse tree'' \\
\textbf{Component-based design} & ``\textbf{specification}'' &
``\textbf{instance}'' \\
\end{tabular}
\caption{Vocabulary for models and instances}\label{tab:modinst}
\end{table}

A reason why this distinction is often not needed or used is that
most systems traditionally have a one-to-one mapping between
specifications and instances. In the example above, in the early age
of the Internet the specification would be reified using exactly one
Squid process, one Apache process, one PHP interpreter and one Oracle
process. Both the application programmer and the system administrator
could then use the same words ``the Apache server'' to designate
either the specification or the instance, using context to
disambiguate meaning.

\section{Coordination \vs computation}
\label{sec:coord}

Beyond the basic definitions of components, component-based design
relies on \emph{compositionality}: defining new aggregate or
\emph{composite} components built out of sub-components.
To achieve this, an application designer works in a \emph{coordination
  environment} which provides both facilities to specify composites,
\ie a \emph{coordination language}, and to run these composite
specifications, \ie a \emph{coordinating run-time system}.

The characteristic of coordination environments is that they can be
fully defined and implemented before the library of actual primitive
components is known. This can be illustrated with the example of
Unix. With Unix, an operating system kernel can be implemented to run
commands from disk before the commands themselves are
implemented. Moreover, both Unix ``shell'' interpreters and scripts
can be implemented and validated, also independently from the commands
they will invoke at run-time.

This separation is possible because \emph{interfaces in component
  specifications map to interfacing mechanisms between component instances}
at run-time. With Unix, command-component interfaces are specified via
their acceptable command-line arguments and how they promise to behave
with regards to network, file, signal and IPC operations. At run-time,
these specifications are mapped to uses of system calls.

A coordination environment is thus composed of:
\begin{itemize}
\item a \emph{run-time system} where execution occurs, which can be extended
  with new components after the system is implemented;
\item a \emph{coordination language}, where a designer can specify
  external primitive components by interface only, and composites
  thereof;
\item an \emph{interfacing mechanism} in the run-time environment, between the coordination system
  and component implementations;
\item \emph{language semantics that guarantee common run-time properties
  over composites}, without requiring a full definition of
  the primitive components (since this definition may not be known at
  specification time).
\end{itemize}

\begin{table}[h]
\centering
\footnotesize
\begin{tabular}{p{.12\textwidth}p{.12\textwidth}p{.2\textwidth}p{.45\textwidth}}
Environment & Coordination language & Specification construct for
external primitive components & Run-time mechanism for interfacing \\
\hline
POSIX & POSIX API & \texttt{fork}/\texttt{exec} & File, network, signal and
IPC system calls \\
GHC and run-time & Haskell & \texttt{foreign} & Any of C/C++, .NET, JVM, Windows or other
system-specific ABI call conventions \\
.NET & C\#, F\#, VB\# etc. & \texttt{DllImport}/\texttt{extern} &
Dynamic linker and standard call convention \\
C & C & \texttt{extern} & Static linker and  ABI call convention \\
C & C & \texttt{asm} & Processor's Instruction Set Artchitecture \\
C/Unix & C & \texttt{dlopen}/\texttt{dlsym} & Dynamic linker and ABI
call convention \\
JVM & Java, Scala & \texttt{native} & Linker and ABI call convention  \\
Common LISP & LISP & \texttt{defctype}, \texttt{defcfun} & Linker and
ABI call convention \\
CPython  & Python & \texttt{import} & Linker and ABI call convention
\\
S-Net and run-time & S-Net & \texttt{box} & Linker and ABI call convention \\
\end{tabular}
\caption{How existing programming environments provide interfaces for
externally defined primitive components}\label{tab:ifs}
\end{table}

This definition partly contrasts with what we can call
``computation'' languages.  The part of a language dedicated to
computation requires that the language semantics fully define how to
manipulate \emph{data values} and how to explicitly express them using
\emph{literals} within the language. However, most programming
languages oriented towards computation are actually implemented using
a coordination environment. For example, C programs can use
\texttt{extern} definitions and the \texttt{asm} statement to compose
behavior from components only known in the execution
environment. Other examples are given in \cref{tab:ifs}.

\begin{figure}
\centering
\includegraphics[width=.8\textwidth]{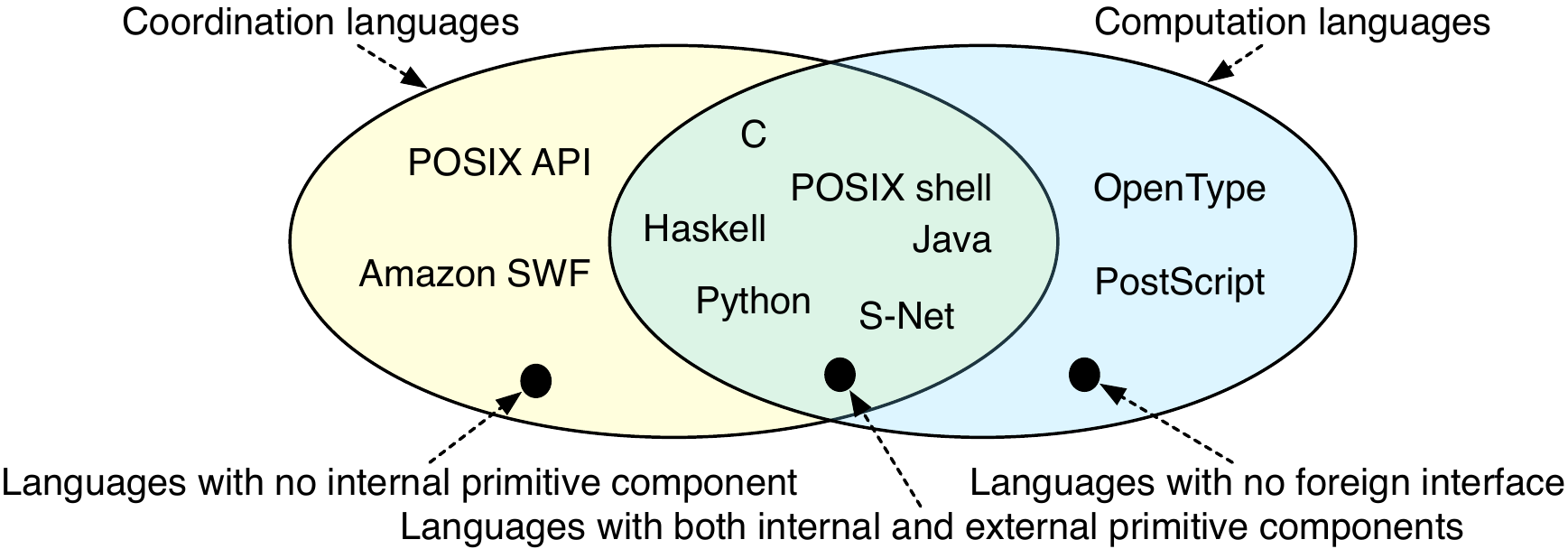}
\caption{Coordination \vs computation: a Venn
  diagram}\label{fig:compcoord}
\end{figure}

In other words, coordination environments are a subset of programming
environments, which \emph{also} contain computation environments. This
relationship is illustrated in \cref{fig:compcoord}.

The availability of a foreign interface depends on the language,
and without it a computation language cannot be used for
coordination. For example, PostScript lacks the ability to define
components externally and is thus ``only'' a computation language.

Complementarily, a language may offer a foreign interface but no
expressivity to express computations ``within it''.  This is the case
of Amazon's Software Workflow
Framework\footnote{\url{https://aws.amazon.com/swf/}}, as of this
writing, which only manipulates services defined externally. This type
of language can thus be called ``purely coordinating.''
Note
that S-Net, which is usually ``marketed'' as a coordination language,
does provide facilities to express literals and compute over them
(record tags and filters) and can thus be used for computation
without using external primitive components at all.

\section{Discussion}
\label{sec:discuss}

Perhaps unsurprisingly, the formulation so far confirms the idea that
most programming languages contain some elements of
coordination. However, it also confirms that coordination can be
recognized using a clear-cut criterion.

This criterion is whether the language designer enables a programmer
to \emph{import and use new primitive components not defined by the
language itself and which may only be fully known in the
  run-time environment}. Using this criterion, one can recognize that
PostScript, for example, does not support coordination.

However, the fact remains that most languages contain both elements of
computation and coordination. Also, the expressive power of a language
focused on coordination can be equally well be constructed using
abstractions within a language focused on computation that also
provides a foreign interface.
The two fundamental questions that motivated this analysis thus remain:
\begin{enumerate}
\item is there room in the technology landscape for languages and
  run-time systems designed and advertised mainly towards coordination?
\item if so, should they be implemented using their own technology
  stack, or instead as libraries of constructed abstractions
  within established languages?
\end{enumerate}

There are two arguments in favor of the first point. One is that
industrial software engineering acknowledges, supports and extensively
exploits black-box design when designing large applications. This
audience may find interest in technology that acknowledges
componentization and promotes willful ignorance of component
definitions while specifying composites. The other argument is
reuse: it is a fact that software components already exist in
different languages, and coordination technology that can integrate
them together enables more reuse.

There are also two arguments against. One is that different languages
create fragmentation of expertise in the population of programmers and
subsequently effort duplication: common coordination features end up
being implemented both as features of coordination languages and as
libraries in computation languages. This translated to redundancy of
human effort. The second is that systematic componentization creates
an artifical barrier to cross-layer optimizations, for example
inter-procedural optimizations in a compiler between the procedures of
different components. In other words, componentization is a likely
source of run-time inefficiency.

From the implementation perspective (the second point mentioned
above), the argument in favor of a separate technology is one of
research efficiency. Indeed, in a research environment, time and effort
are precious. The luxury of a specialized implementation enables
researchers to focus on issues specific to coordination, without
requiring them to care about integration in a more general language
whose implementation is thus also necessarily more complex.

The argument against is, again, fragmentation of expertise. When
embedding coordination in an existing language substrate, research
projects can recruit new members from the pool of existing programmers
acquainted with that language's implementations. If the substrate
language and implementations are different/new, recruiting is more
difficult and/or implies more training overhead.

The discussion about the consequence of these trade-offs on the future
of S-Net and related work lies outside the scope of this report.

\section{Conclusion}

This report has confirmed the existence of a view in which
coordination languages and programming languages are not two disjoint
sets. Programming for coordination can be seen as a mere \emph{style}
of programming, which requires a \emph{technical} means by which
composite software applications can be specified using components that
are only available separately, or later in time. This technical means
is the availability of a foreign component interface in the language,
and we propose to call ``coordination languages'' those language who
provide this facility.

Meanwhile, most currently used languages provide foreign interfaces,
and thus exhibit some elements of coordination. However not all do,
and the availability of a foreign interface thus forms an objective
and qualitative criterion to identify languages that can be used for
coordination.

However, despite the availability of a clear-cut qualitative
definition of coordination, the segregation of academic attention away
from general language design and implementation has non-obvious
cost/benefit trade-offs, mostly related to duplication of effort and
skill fragmentation across language boundaries.

\section*{Acknowledgements}
\addcontentsline{toc}{section}{Acknowledgements}

This document reports on thoughts nurtured during successive
discussions with Merijn Verstraaten, Alex Shafarenko, Sven-Bodo
Scholz, Kath Knobe, Roy Bakker, Michiel W. van Tol and Sebastian
Altmeyer.

\newcommand{\etalchar}[1]{#1} % pre-defined so the bst does not complain
\addcontentsline{toc}{section}{References}
\bibliographystyle{is-plainurl}
\bibliography{literature}

\end{document}